\documentstyle[12pt]{article}
\begin{document}
\begin{center}
{\large { MAXWELL-CHERN-SIMONS THEORY IS FREE FOR MARGINALLY NONCOMMUTATIVE SPACETIMES}}
\vskip 2cm
Subir Ghosh\\
\vskip 1cm
Physics and Applied Mathematics Unit,\\
Indian Statistical Institute,\\
203 B. T. Road, Calcutta 700108, \\
India.

\vskip 3cm
 Abstract:\\
 \end{center}
 We have conclusively established the duality between noncommutative
 Maxwell-Chern-Simons theory and Self-Dual model, the latter in ordinary spacetime,
 to the first non-trivial order in the noncommutativity parameter $\theta^{\mu\nu}$,
 with $\theta^{0i}=0$. This shows that the former theory is free for marginally
 noncommutative spacetimes. A $\theta$-generalized
 covariant mapping between the variables of the two models in question
 has been derived explicitly, that converts one model to the other,
 including the symplectic structure and action.


\newpage

In this Letter we provide an example of a Non-Commutative (NC)
{\it{free}} field theory in 2+1-dimensions - the NC abelian
Maxwell-Chern-Simons (MCS) theory. Our analysis is perturbative in
$\theta^{\mu\nu}$ - the noncommutativity parameter - to the first
nontrivial order. Hence the result is valid for spacetimes with
small noncommutativity. This is a non-trivial result since the
noncommutativity generates non-linear derivative type of
interaction terms in the action. We show in a conclusive way that
NC MCS theory is dual to the abelian Self-Dual (SD) model (in
{\it{ordinary}} spacetime). The latter model was shown to
represent a free massive spin one excitation by Deser and Jackiw
\cite{dj}. They also proved that SD model was dual to the
well-known MCS (topologically massive gauge) theory
\cite{djt}{\footnote{It is important to note that the effect of
the Chern-Simons term is not perturbative in nature. In its
absence, the 2+1-dimensional pure Maxwell theory describes a free
massless spin zero  excitation \cite{djt}.}. The importance of the
SD model was further enhanced when it was shown to appear in the
bosonization \cite{fs} of the fermionic massive Thirring model in
the large fermion mass limit. In a generic way, the planar gauge
theories have played important roles in the context of physically
interesting phenomena (that are effectively planar), such as
quantum Hall effect, high-$T_C$ superconductivity, to name a few
and in anyon physics, where excitations having arbitrary spin and
statistics appear.

We restrict ourselves to only spatial noncommutativity
($\theta^{0i}=0$) and the results are valid to the first
non-trivial order in $\theta^{\mu\nu}$ - the noncommutativity
parameter, as defined below,
\begin{equation}
[x^{\rho},x^{\sigma}]_{*}=i\theta^{\rho\sigma}.
\label{nc}
\end{equation}
The $*$-product is given by the Moyal-Weyl formula,
\begin{equation}
p(x)*q(x)=pq+\frac{i}{2}\theta^{\rho\sigma}\partial_{\rho}p\partial_{\sigma}q+~O(\theta^{2}).
\label{mw}
\end{equation}
The reason for invoking $\theta^{0i}=0$ is that space-time
noncommutativity can induce higher order time derivatives leading to a
loss of causality. Also, even to $O(\theta )$, it can alter the symplectic
structure in a significant way, that might result in a non-perturbative change
in the dynamics, which we want to avoid.

The study of NC quantum field theory has acquired a prominent
place after the seminal work of Seiberg and Witten \cite{sw}, who
showed that NC manifolds emerge naturally in the context of
$D$-branes on which an open string can terminate, (in the presence
of a two-form background field). Field theories living on
$D$-branes are essentially NC. The effects of noncommutativity can
be systematically studied in a perturbative way by exploiting the
Seiberg-Witten map  \cite{sw}, which converts an NC theory to a
conventional theory in ordinary spacetime. The NC effects appear
as interaction terms. Even though the basic field theoretic
framework remains unaltered (for $\theta^{0i}=0$), the
noncommutativity induces a plethora of distinctive features, such
as UV/IR mixing \cite{1}, presence of solitons in higher
dimensional scalar theory \cite{2}, excitations of dipolar nature
\cite{3}, to name a few. Thus one is interested to know how the
noncommutativity affects established properties of conventional
field theories and one such area is the duality (or equivalence)
between field theories - in particular the MCS-SD duality
\cite{dj}, the case of present interest. It has been shown
\cite{jabbar} that the NC Chern-Simons theory is free. However, in
our approach of exploiting the Seiberg-Witten map, this result is
expected since under the above mapping, the NC Chern-Simons theory
reduces to comutative Chern-Simons theory to all orders of
$\theta$ and hence the results corresponding to commutative
Chern-Simons theory should hold.

It is worthwhile to point out the rationale of restricting the
analysis to $O(\theta)$ only in the present context. Ever since
the advent of noncommutativity feature in spacetime, $O(\theta)$
results in noncommutative field theory have played a significant
role, primarily because in many cases, $O(\theta)$ modifications
tend to depart from commutative spacetime results in a nontrivial
way. As we are going to study the effects of NC spacetime on an
already established result in ordinary spacetime, ({\it{i.e.}} the
MCS and Self-Dual model equivalence \cite{djt}), it is indeed
logical to look for modifications in the leading order in
$\theta$. This is quite in keeping with the spirit of our  work
\cite{sgg,sgg1} where effects of noncommutativity  are looked at
in the context of $CP(1)$ solitons \cite{cp} and in solitons in
the Chern-Simons-Higgs system \cite{csh} respectivly. On the other
hand, as a contrasting example, one should consider the case of NC
solitons in a scalar theory in \cite{2}, where a {\it {large}}
$\theta$  limit has been pursued. This is natural since these
latter NC solitons are completely new entities of NC spacetime,
having no counterpart in the corresponding ordinary spacetime
theory. Furthermore, our methodology relies heavily on the
Seiberg-Witten map \cite{sw}, which is free from any ambiguity up
to $O(\theta)$.

Recently several papers have appeared in this context \cite{sg1}
\cite{of} \cite{bot} \cite{wot} and the results have not always
agreed. The primary reason for the ambiguity is that all  the
works have tried to establish the duality by way of comparing the
actions and showing that they are related in some way \cite{of}
\cite{bot}, such as via a "Master" Lagrangian  \cite{dj}. In
\cite{wot}, two actions are termed as dual when one of them
becomes equal to the other on the surface of the equations of
motion. But in both of the above instances, it is not clear
whether the two actions, satisfying {\it{only}} the above criteria
of duality, share the same symplectic structure and subsequent
dynamics.

We point out that no discussions on the symplectic structure of
the variables (that dictates the dynamics) or an explicit mapping
between the degrees of freedom of the two purported dual theories,
 (NC MCS and NC SD), have been attempted so far. By itself, relating the
actions can not prove duality conclusively and should only be
considered as a confirmatory test of duality, obtained in a  more
fundamental way, concerning the basic fields. The latter scheme
obviously suggests the former relation  at the level of actions.
In fact it should be remembered that in the original work
\cite{dj}, the duality was first proved at the level of symplectic
structures of MCS and SD models. Subsequently a mapping was
provided which can bodily convert  one model in to the other
totally and only then a Master action was provided to corroborate
the previous findings. We will precisely follow this route but
will not attempt the last one - construction of the Master action.

Our metric is $g^{\mu\nu}=diag(1,-1,-1)$. The NC MCS model is
defined in the following way \cite{of},
\begin{equation}
\hat {\cal {L}}_{MCS}=\int d^{3}x[-\frac{1}{4}\hat F^{\mu\nu}*\hat
F_{\mu\nu}+ \frac{m}{2}\epsilon^{\mu\nu\lambda}(\hat
A_{\mu}*\partial_{\nu}\hat A_{\lambda}+\frac{2}{3}i\hat
A_{\mu}*\hat A_{\nu}*\hat A_{\lambda})], \label{lmcs}
\end{equation}
where
$$\hat F_{\mu\nu}=\partial_{\mu}\hat A_{\nu}-\partial_{\nu}\hat A_{\mu}-i\hat A_{\mu}*\hat A_{\nu}+i\hat A_{\nu}*\hat A_{\mu}. $$
We use the NC extension of the CS action derived in \cite{gs}.
Utilizing the Seiberg-Witten map, to the lowest non-trivial order
in $\theta$,
\begin{equation}
\hat A_{\mu}=A_{\mu}+\theta^{\sigma\rho}A_{\rho}(\partial_{\sigma} A_{\mu}-\frac{1}{2}\partial_{\mu} A_{\sigma})~;~
\hat F_{\mu\nu}=F_{\mu\nu}+\theta^{\rho\sigma}(F_{\mu\rho}F_{\nu\sigma}-A_{\rho}\partial_{\sigma} F_{\mu\nu}),
\label{swm}
\end{equation}
we arrive at the following  $O(\theta)$ modified form of the NC
MCS theory, expressed in terms of ordinary spacetime variables
\begin{equation}
\hat {\cal {A}}_{MCS}=\int
d^{3}x[-\frac{1}{4}(F^{\mu\nu}F_{\mu\nu}+2\theta^{\rho\sigma}(F^{\mu}_{~\rho}F^{\nu}_{~\sigma}F_{\mu\nu}
-\frac{1}{4}F_{\rho\sigma}F^{\mu\nu}F_{\mu\nu})) +
\frac{m}{2}\epsilon^{\mu\nu\lambda}A_{\mu}\partial_{\nu}A_{\lambda}],
\label{mcs}
\end{equation}
where $F_{\mu\nu}=\partial_{\mu}A_{\nu}-\partial_{\nu}A_{\mu}$. It
should be remembered that under the Seiberg-Witten map, the NC CS
term exactly reduces to the CS term in ordinary spacetime. In
2+1-dimensions, the action (\ref{mcs}) is further simplified to,
\begin{equation}
\hat {\cal {A}}_{MCS}=\int
d^{3}x[-\frac{1}{4}F^{\mu\nu}F_{\mu\nu}+
\frac{m}{2}\epsilon^{\mu\nu\lambda}A_{\mu}\partial_{\nu}A_{\lambda}
 -\frac{1}{8}\theta^{\rho\sigma}F_{\rho\sigma}F^{\mu\nu}F_{\mu\nu}],
\label{mmcs}
\end{equation}
 and leads to the equation of motion,
\begin{equation}
\partial_{\mu}F^{\mu\nu}+\frac{m}{2}\epsilon^{\nu\alpha\beta}F_{\alpha\beta}+\frac{1}{4}\partial_{\mu}(\theta^{\mu\nu}F^{2}+2(\theta .F)F^{\mu\nu})=0.
\label{eqa}
\end{equation}
As we are interested in the symplectic structure, we now move on
to the Hamiltonian formulation of the above model and introduce
the canonical momenta and Poisson brackets,
\begin{equation}
\Pi_{\mu}\equiv \frac{\delta \hat{\cal{L}}_{MCS}}{\delta \dot A^{\mu}}~;~
\{A^{\mu}(x,t),\Pi^{\nu}(y,t)\}=-g^{\mu\nu}\delta (x-y).
\label{pb}
\end{equation}
Here the non-trivial momenta are
\begin{equation}
\Pi_{i}=(1-\theta B)F^{0i}+\frac{m}{2}\epsilon^{ij}A^{j}.
\label{mom}
\end{equation}
Reverting to a non-covariant notation \cite{dj},
$$E^{i}=F^{i0},~B=-\epsilon^{ij}\partial^{i}A^{j}=-F^{12}$$
the momenta and Hamiltonian are obtained as,
\begin{equation}
\Pi_{i}=(1-\theta B)E_{i}-\frac{m}{2}\epsilon_{ij}A_{j},
\label{cmom}
\end{equation}
$$
\hat{\cal{H}}_{MCS}\equiv \Pi_{i}\dot A^{i}-\hat{\cal{L}}_{MCS}=\frac{1}{2}(1-\theta B)[(1-\theta B)^{-2}(\Pi^{i}+\frac{m}{2}\epsilon^{ij}A^j)^{2}+B^{2}]$$
$$
+A^{0}[\partial^{i}(\Pi^i+\frac{m}{2}\epsilon^{ij}A^j)+mB]$$
\begin{equation}
=\frac{1}{2}(1-\theta B)(E^iE^i+B^2)+A^0\hat{\cal{G}},
\label{ham}
\end{equation}
where the Gauss law constraint appears as
\begin{equation}
\hat{\cal{G}}\equiv \partial^{i}[(1-\theta B)E^i]+mB \approx 0.
\label{gau}
\end{equation}
The relation (\ref{e}) has been used to derive (\ref{ham}).
Inverting the relation (\ref{mom}) to express the electric field
in terms of phase space variables,
\begin{equation}
E^i=(1-\theta B)^{-1}(\Pi^{i}+\frac{m}{2}\epsilon^{ij}A^j)\approx (1+\theta B)(\Pi^{i}+\frac{m}{2}\epsilon^{ij}A^j),
\label{e}
\end{equation}
it is straightforward to compute the following algebra among the
electric and magnetic fields,
$$
\{E^{i}(x),E^{j}(y)\}=m\epsilon^{ij}(1+2\theta B)\delta (x-y)-\theta [\epsilon^{kj}E^i(x)+\epsilon^{ki}E^j(y)]\partial^{k}_{(x)}\delta (x-y),$$
\begin{equation}
\{E^{i}(x),B(y)\}=\epsilon^{ij}[1+\theta B(x)]\partial^{j}_{(x)}\delta (x-y)~;~~\{B(x),B(y)\}=0.
\label{alg}
\end{equation}
Interestingly, similar to the $\theta =0$ theory \cite{dj}, there
exists a free field representation of $E^i$ and $B$ in terms of
$(\varphi ,\pi )$ obeying $\{\varphi (x),\pi (y)\}=\delta (x-y)$,
\begin{equation}
B\equiv \sqrt{-\nabla^{2}}\varphi ~,~E^i\equiv (1+\theta B)(\epsilon^{ij}\hat \partial_{j}\pi-m\hat \partial_{i}\varphi ),
\label{can}
\end{equation}
that satisfies the algebra (\ref{alg}). The notations used are
$\nabla^{2}\equiv \partial^{i}\partial^{i},~\hat \partial^{i}\equiv \frac{\partial^{i}}{\sqrt{-\nabla^{2}}}$.

 The backbone of our subsequent analysis is the crucial observation that a new set of variables $(\tilde E^i,~\tilde B)$ can be introduced,
\begin{equation}
\tilde E^{i}\equiv(1-\theta B)E^i~,~\tilde B\equiv B,
\label{new}
\end{equation}
that obeys the $\theta =0$ algebra,
$$
\{\tilde E^{i}(x),\tilde E^{j}(y)\}=m\epsilon^{ij}\delta (x-y),$$
\begin{equation}
\{\tilde E^{i}(x),\tilde B(y)\}=\epsilon^{ij}\partial^{j}_{(x)}\delta (x-y)~;~~\{\tilde B(x),\tilde B(y)\}=0.
\label{alg1}
\end{equation}
Exploiting the inverse relations,
\begin{equation}
E^i=(1-\theta B)^{-1}\tilde E^i\approx(1+\theta \tilde B)\tilde E^i~;~~B=\tilde B,
\label{old}
\end{equation}
the Gauss law constraint and the Hamiltonian is rewritten below,
\begin{equation}
\partial^{i}\tilde E^i+m\tilde B\approx 0,
\label{newg}
\end{equation}
\begin{equation}
{\cal{H}}_{MCS}=\frac{1}{2}[\tilde E^{i}\tilde E^{i}+\tilde B^2+\theta \tilde B(\tilde E^{i}\tilde E^{i}-\tilde B^2)].
\label{nh}
\end{equation}
Since the $(\tilde E^i,~\tilde B)$ algebra has become
$\theta$-independent, we can use the ordinary spacetime free field
representation,
\begin{equation}
\tilde B\equiv \sqrt{-\nabla^{2}}\varphi ~,~\tilde E^i\equiv \epsilon^{ij}\hat \partial_{j}\pi-m\hat \partial_{i}\varphi .
\label{can1}
\end{equation}
Indeed, the theory appears to be far from being free since  the
$\theta$-contribution in the Hamiltonian (\ref{nh}) has apparently
turned the theory in to a non-local one. This is because unlike
the $\theta =0$ part, which is quadratic, the $\theta$-term is of
higher order and the non-local operators involved in the free
field representation (\ref{can1}) can not be shifted around, even
under the integral. However, we now show that to $O(\theta)$, this
theory can be identified to the abelian Self Dual theory {\it{in
ordinary spacetime}} by means of a Lorentz covariant mapping of
the degrees of freedom. This constitutes our main result. The SD
theory was solved long time ago \cite{dj}. It represents a single
free massive spin one mode.

The all important mapping between NC MCS variables $(\tilde E^i,\tilde B)$ and SD variables $(f^{\mu})$ is
\begin{equation}
\tilde E^i\equiv \epsilon^{ij}f^j ~,~\tilde B\equiv -(f^0+\theta X),
\label{map}
\end{equation}
which has the covariant structure,
\begin{equation}
\frac{1}{2}\epsilon^{\mu\nu\lambda}\tilde F_{\nu\lambda}\equiv f^\mu+\frac{1}{2}\epsilon^{\mu\nu\lambda}\theta_{\nu\lambda}X .
\label{map1}
\end{equation}
$X$ is an as yet unknown scalar variable. Note that the $\theta
=0$ mapping was first given in \cite{dj}. The NC extension of the
map is such that {\it{only}} the identification between  the
non-dynamical time-components of the respective vector fields is
affected.

We now directly exploit
(\ref{map}) to express ${\cal{H}}_{MCS}$ in (\ref{nh}) in terms of
$f^\mu$ variables,
\begin{equation}
{\cal{H}}_{f}=\frac{1}{2}[f^if^i+f^0f^0+2\theta f^0X+\theta f^0f^\sigma f_\sigma].
\label{sd}
\end{equation}
This is trivially obtained from the first order Lagrangian,
\begin{equation} {\cal{L}}_f=\frac{1}{2}f^{\mu}f_{\mu}-\frac{1}{2m}\epsilon^{\mu\nu\lambda}f_{\mu}\partial_{\nu}f_{\lambda} -\frac{1}{4}\epsilon^{\mu\nu\lambda}\theta_{\nu\lambda}f_{\mu}f^{\sigma}f_{\sigma}-\frac{1}{2}\epsilon^{\mu\nu\lambda}\theta_{\nu\lambda}f_{\mu}X.
\label{lsd}
\end{equation}
The $\theta$-term in (\ref{lsd}) can be removed by putting
$X=-\frac{1}{2}f^\sigma f_\sigma $ and we are left with the
abelian Self Dual Model in ordinary spacetime,
\begin{equation}
{\cal{L}}_{SD}=\frac{1}{2}f^{\mu}f_{\mu}-
\frac{1}{2m}\epsilon^{\mu\nu\lambda}f_{\mu}\partial_{\nu}f_{\lambda}.
\label{lsd1}
\end{equation}
This is our cherished result.

Obviously this is the most economical form of the dual theory.
Some amount of non-uniqueness creeps in through a non-vanishing
$X$. However, as we have emphasized, $X$ has to be such that it
does not vitiate the symplectic  structure between $f^i$, the
independent dynamical degrees of freedom.

Some comments of the related recent works in the perspective of
the present analysis is in order. The bone of contention happens
to be the NC generalization of the SD model. We have argued before
(Ghosh in \cite{sg1}, \cite{sg2}, \cite{sub}) that since the SD
theory is a quadratic theory with no gauge invariance, its natural
extension in the NC regime should be the same as the original
theory. This has also been demonstrated in the context of NC
Soldering phenomena \cite{sub} (see \cite{st} for reviews on
Soldering formalism). This idea is echoed in the present work as well.

Next we come to the work in  \cite{bot} where the duality between
NC MCS and NC SD model was studied from the Master action point of
view, where the NC SD model contains the NC Chern-Simons term. At
first sight it seems that this observation can be accommodated in
our analysis, since for $\theta^{0i}=0$, the extra $\theta
$-dependant three $f^{\mu}$ term
$\epsilon^{\mu\nu\lambda}\theta^{\alpha\beta}f_{\mu}\partial_{\alpha}f_{\nu}\partial_{\beta}f_{\lambda}$
in the NC SD action of \cite{bot} reduces to
$\epsilon^{ij}\theta^{kl}f_0\partial_{k}f_{i}\partial_{l}f_{j}$
modulo total derivatives. Note that this term will not affect the
symplectic structure of the SD model but will modify the
constraint connecting $f^0$ to $f^i$. But the problem is that this
action can not be generated from NC MCS theory by any covariant
mapping between $F^\mu$ variables (of NC MCS) and $f^\mu$
variables (of NC SD) as in (\ref{map1}), without modifying the
$\{f^i,f^j\}$ symplectic structure, that governs the dynamics. The
speciality of the mapping (\ref{map1}) is that it keeps the
$F^i\Leftrightarrow f^i$ identification unaltered. In this sense the duality
derived in \cite{bot} is weaker because the covariance in the
mapping will be lost.

On the other hand, \cite{wot} has started from the NC SD model
(with the NC Chern-Simons term) and has obtained a dual theory
which differs from the NC MCS theory. It will be interesting to
redo the analysis along the lines demonstrated here taking the
particular version of NC SD theory in \cite{wot} as the starting
point.

As a final remark, we mention that our analysis of the duality
relation is classical in nature and so we were able to exploit the
classical result that the NC CS theory is mapped to the ordinary
CS theory via the Seiberg-Witten map. Obviously, the
non-perturbative effects of $\theta$ leading to the quantization
of the level of the NC CS model \cite{nair} in the quantum theory
can not be addressed in the perturbative framework of the
Seiberg-Witten map. In fact, even in a perturbative computational
scheme, recently it has been shown \cite{oog} that mapping between
NC and ordinary CS theories as quantum theories, requires a
modification in the Seiberg-Witten map itself, where quantum
corrections are to be incorporated. As an interesting future
problem, one might study the quantum equivalence between the NC
and ordinary Self-Dual models along the lines of \cite{oog} since
in this case also, both ordinary and NC forms of the actions are
known exactly. This is essential for the perturbative analysis
\cite{okawa}. This will also establish the duality between NC MCS
and NC Self-Dual models in their quantized version, thus
generalizing the $O(\theta)$ classical result presented here.
\vskip.5cm {\bf {Acknowledgements:}} It is a pleasure to thank
Marcelo Botta Cantcheff, Pablo Minces and Clovis Wotzasek for
correspondence.


\vskip 2cm
\begin{thebibliography}{99}
\bibitem{dj}S.Deser and R.Jackiw, Phys.Lett. 139B(1984)371.
\bibitem{djt}S.Deser, R.Jackiw and S.Templeton, Phys.Rev.Lett. 48 (1982)975; Ann.Phys. 140 (1982)372.
\bibitem{fs}E.Fradkin and F.A.Schaposnik, Phys.Lett. 338B(1994)253; R.Banerjee, Phys.Lett. B358(1995)297.
\bibitem{sw}N.Seiberg and E.Witten, JHEP 9909(1999)032. For reviews, see for example M.R.Douglas and N.Nekrasov, Rev.Mod.Phys. 73 (2001)977; R.J.Szabo, Phys.Rept. 378 (2003)207.
\bibitem{1}S.Minwalla, M. Van Raamsdonk and N.Seiberg, {\it {Noncommutative perturbative dynamics}}, hep-th/9912072.
\bibitem{2}R.Gopakumar, S.Minwalla and A.Strominger, JHEP 0005 (2000)020.
\bibitem{3}M.M.Sheikh-Jabbari, Phys.Lett. B455 (1999)129; D.Bigatti and L.Susskind, Phys.Rev.
D62 (2000) 066004; S.Ghosh, Phys.Lett. B571 (2003)97.
\bibitem{jabbar}A.Das and M.M.Sheikh-Jabbari, JHEP 0106 (2001)028.
\bibitem{sgg} S.Ghosh, Nucl.Phys. B670 (2003)359 (hep-th/0306045); hep-th/0310155.
\bibitem{sgg1}S.Ghosh, in preparation.
\bibitem{cp} F.Wilczek and A.Zee, Phys.Rev.Lett. 51 (1983)2250.
\bibitem{csh}J.Hong, Y.Kim and P.Y.Pac, Phys.Rev.Lett. 64 (1990)2230; R.Jackiw and E.J.Weinberg,  Phys.Rev.Lett. 64 (1990)2234.
\bibitem{sg1}S.Ghosh, Phys.Lett. B558 (2003)245 (hep-th/0210107); M.Botta Cantcheff and P.Minces, Phys.Lett. B557 (2003)283 (hep-th/0212031).
\bibitem{of}O.F.Dayi, Phys.Lett. B560 (2003)239.
\bibitem{bot}M.Botta Cantcheff and P.Minces, {\it {Duality in Noncommutative Topologically Massive Gauge Field Theory Revisited}}, hep-th/0306206.
\bibitem{wot}T.Mariz, R. Menezes, J.R.S. Nascimento, R.F.Ribeiro, C. Wotzasek, {\it{Duality on noncommutative manifolds: the non-equivalence between self-dual and topologically massive models}}, hep-th/0306265.
\bibitem{gs}N.Grandi and G.A.Silva, Phys.Lett. B507 (2001)345.
\bibitem{sg2}S.Ghosh, Phys.Lett. B563(2003)112.
\bibitem{sub}S.Ghosh, Phys.Lett. B579 (2004)377 (hep-th/0306273).
\bibitem{st}M.Stone, University of Illinois Preprint, ILL-TH-28-89; E.M.C.Abreu, R.Banerjee and C.Wotzasek, Nucl.Phys. B509(1998)519. For a recent review and references, see C.Wotzasek, {\it{Duality and Topological Mass Generation in Diverse Dimensions}}, hep-th/0305127.
\bibitem{nair} V.P.Nair and A.P.Polychronakos, Phys.Rev.Lett. 87 (2001)030403 (hep-th/0102181); D.Bak, K.M.Lee and J.H.Park, Phys.Rev.Lett. 87 (2001)030402 (hep-th/0102188).
\bibitem{oog}K.Kaminsky, Y.Okawa and H.Ooguri, Nucl.Phys. B663 (2003)33 (hep-th/0301133).
\bibitem{okawa}Y.Okawa, private communication.

\end{thebibliography}
\end{document}